\documentclass[twocolumn,trackchanges]{aastex631}

\usepackage{xcolor}       
\usepackage{soul}
\usepackage{ulem}

\begin{document}

\graphicspath{{./}{figures/}}


\shorttitle{Dust from Splash}
\shortauthors{Becker et al.}

\title{Releasing Atmospheric Martian Dust in Sand Grain Impacts}


\author{T. Becker}
\affiliation{University of Duisburg-Essen, Faculty of Physics, Lotharstr. 1, 47057 Duisburg, Germany}

\author{J. Teiser}
\affiliation{University of Duisburg-Essen, Faculty of Physics, Lotharstr. 1, 47057 Duisburg, Germany}

\author{T. Jardiel}
\affiliation{Instituto de Cer\'{a}mica y Vidrio  C/Kelsen 5, Campus Cantoblanco 28049 Madrid, Spain}

\author{M. Peiteado}
\affiliation{Instituto de Cer\'{a}mica y Vidrio C/Kelsen 5, Campus Cantoblanco 28049 Madrid, Spain}

\author{O. Munoz}
\affiliation{Instituto de Astrof\'{i}sica de Andaluc\'{i}a, CSIC Glorieta de la Astronom\'{i}a s/n 18008 Granada, Spain}

\author{J. Martikainen}
\affiliation{Instituto de Astrof\'{i}sica de Andaluc\'{i}a, CSIC Glorieta de la Astronom\'{i}a s/n 18008 Granada, Spain}

\author{J.C. Gomez Martin}
\affiliation{Instituto de Astrof\'{i}sica de Andaluc\'{i}a, CSIC Glorieta de la Astronom\'{i}a s/n 18008 Granada, Spain}

\author{G. Wurm}
\affiliation{University of Duisburg-Essen, Faculty of Physics, Lotharstr. 1, 47057 Duisburg, Germany}


\begin{abstract}
Emission of dust up to a few micrometer in size by impacts of sand grains during saltation is thought to be one source of dust within the Martian atmosphere. To study this dust fraction, we carried out laboratory impact experiments. Small numbers of particles of about 200\textmu{}m in diameter impacted a simulated Martian soil (bimodal \textit{Mars Global Simulant}). Impacts occurred at angles of $\sim 18^\circ$ in vacuum with an impact speed of 
$\sim 1 \rm m/s$. Ejected dust was captured on adjacent microscope slides and the emitted particle size distribution (PSD) was found to be related to the soil PSD. We find that the ejection of clay sized dust gets increasingly harder the smaller these grains are. However, in spite of strong cohesive forces, individual impacts emit dust of 1\textmu{}m and less, i.e. dust in the size range that can be suspended in the Martian atmosphere. More generally, the probability of ejecting dust of a given size can be characterized by a power law in the size range between 0.5\textmu{}m and 5\textmu{}m (diameter).
\end{abstract}




\section{Introduction}

Mars has a non-negligible but thin atmosphere with an average surface pressure of about 6 hPa \citep{Zurek2017}. While this is too little to suspend grains as large as Earth's atmosphere is capable of, small dust particles can persist. Airborne dust has been observed and modeled to be on the order of 1\textmu{}m in effective radius during clear skies and temporarily up to a few \textmu{}m during dust storms \citep{Clancy1995, Clancy2010, Maeaettaenen2013, Lemmon2019, Chenchen2019, Daversa2022}. 

Micrometer dust in the atmosphere is a fundamental component of the planet's climate system e.g. increasing the water content at high altitude \citep{Neary2020, Vandaele2019, Holmes2021}. However, dust entrained by the atmosphere is not suspended permanently. There is a continuous sedimentation of dust as evident in the degradation of solar panels on Martian landers \citep{Lorenz2021}. Therefore, atmospheric dust has to be replenished continuously.

With dust devils and local dust storms being observed \citep{Stanzel2008, Reiss2016, Elsaid2020, Baker2021, Favaro2021}, gusts of wind picking up dust at rover sites \citep{Newman2021}, and dunes migrating in some places \citep{Banks2018}, it is obvious that at least some fraction of the entrained dust is related to wind driven activities. Plenty of work has been carried out on wind-driven saltation, where sand is moving by means of short-trajectory hops along the surface, aiding further lifting with each impact \citep{Greeley1976, Greeley1980, Iversen1976, Iversen1982, Merrison2012, kok2014, Musiolik2018, Swann2020, Kruss2020}. This implies a kind of bimodal transport behaviour. While micrometer dust can be injected into the atmosphere, typically, sand grains of about 100\textmu{}m are the easiest to be moved by wind on the surface. 

This existence of a preferred particle size in motion is a consequence of the underlying force balance between the acting forces - gravity, cohesion, and gas drag. For large particles, gravity, going with the volume of a particle, dominates over cohesion which itself linearly depends on the grain size. Therefore, at the large particle size end, gas drag (depending on the particle's cross section) decreases compared to gravity and large grains are ever harder to be moved.

On the small size end, cohesion dominates, eventually, and the ratio between gas drag force and cohesion decreases towards smaller grain size, also making it ever harder to lift smaller grains. Interpolating between both extremes, there is a sweet spot for particles being most susceptible to wind drag at about 100 \textmu{}m under Martian conditions \citep{Iversen1976, Greeley1980, Shao2000}.  
A corresponding plot, in which different measurements have been combined by \citet{Neakrase2016}, shows the threshold friction velocity corresponding to grain size (fig. \ref{Threshold}).

These findings are consistent with observations on Mars. For Gale crater, \citet{Weitz2018} find that, volumentrically, most particles in aeolian bedforms are in the size range of $50 \rm \mu m$ to $150  \rm \mu m$. The grains found by the Phoenix lander are in the range between $20 \rm \mu m$ to $100  \rm \mu m$ \citep{Goetz2010}. Noting that \citet{Goetz2010} give number size distributions, these values are comparable. We considered these size constraints to set up our simulated Martian soil.

Having listed a number of different size fractions by now, we note that not all of these sizes fall in single well defined classification schemes of sediments. The large grains observed on Mars mentioned above span a range from medium silt to fine sand.  Also "dust" is rather unspecified with respect to size so far. To make this somewhat more quantitative and in view of the application to particles within the Martian atmosphere, we refer to "dust" as particles which can be suspended for long time in the atmosphere, i.e. essentially being restricted to clay size particles up to a few micrometer on Mars, reaching slightly into the very fine silt region.

Dust can be lifted during aeolian activity by different mechanisms, with the simplest case being direct dust suspension due to gas drag. For Earth, direct dust entrainment by wind has e.g. been discussed by \cite{Loosmore2000}. However, if potential saltators are present they are picked up first. Martian friction velocities are always just on the lower limit of what is needed to move these saltators \citep{Neakrase2016}, implying they are regularly much too low to pick up dust directly.  
Wind driven dust fluxes are therefore rather regularly observed in combination with saltation of larger sand grains \citep{Shao1993}. 

\begin{figure*}
	\includegraphics[width=\textwidth]{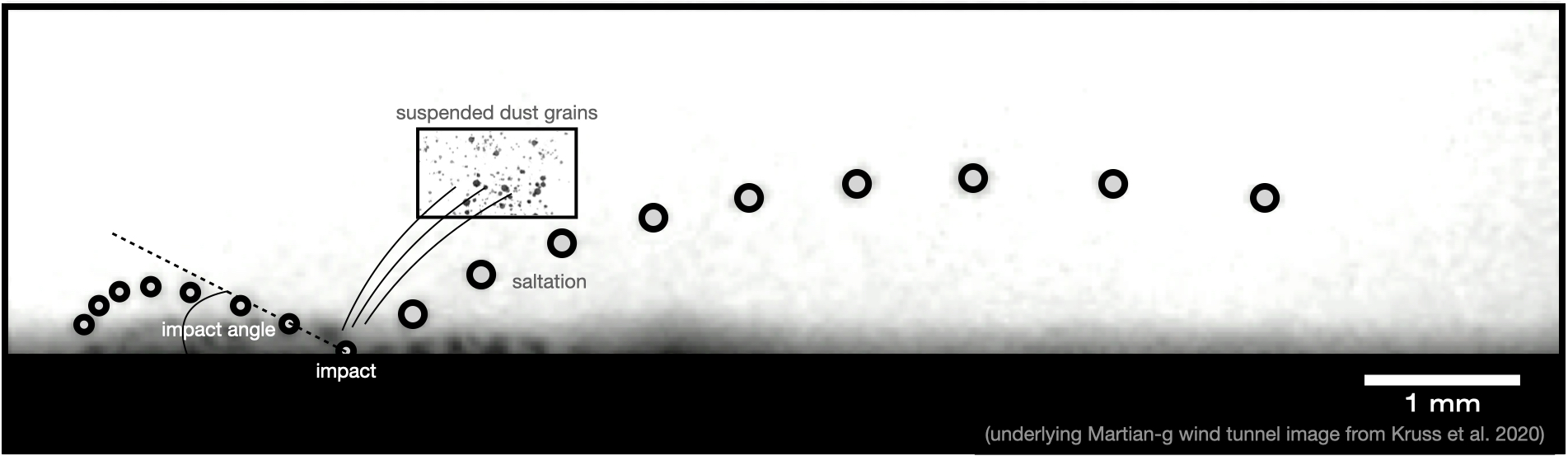}
	\caption{\label{fig.sketch}Principle of dust suspension by impact splash of saltating grains. The underlying image is from \citet{Kruss2020} and shows a superposition of images of saltating grains under Martian gravity with a time interval of about 2 ms. The "suspended dust grains" inset is only a sketch but refers to the dust fraction sampled in the experiments.}
\end{figure*}

With the 100 \textmu{}m fraction being moved first (fig. \ref{Threshold}). it is only natural that wind driven dust flux can only be observed together with a movement of larger sand grains. As those are too heavy to be picked up by the wind they move along ballistic trajectories by steadily bouncing off and reimpacting the surface after being liberated by the wind.
This idea is visualized in fig. \ref{fig.sketch}, which, in principle, is not new but similar to many standard sketches of this kind. In our choice of sketch, we use an image from a wind tunnel experiment by \citet{Kruss2020} as underlying background. We do so in order to point out the major aspects for the experiments reported here. The wind tunnel experiment was carried out at the threshold of saltation under Martian gravity on a parabolic flight. The size of the observed saltating grains was on the order of 100\textmu{}m. 
Furthermore, impact velocities at the threshold are of the order of 1 m/s and impacts occur inclined. Therefore, this sketch already includes some quantitative aspects. However, the dust fraction being present in these experiments could not be detected. 

Additional to direct dust entrainment or liberation of dust by saltating impacts of sand grains, other mechanisms are deemed to have an effect on ejecting dust into the Martian atmosphere, as well. Electrostatic forces are among those mechanisms. It has e.g. been shown that electrostatic forces can directly aid lifting \citep{Renno2008, Holstein2010}. \citet{Kok2008} showed, that saltating particles can charge by repeated bouncing from the surface. Those charged particles then build an electrical field between them and the ground \citep{Schmidt1998, Zheng2003}. This further aids lifting of charged dust grains emitted upon saltating impacts. In fact, field measurements by \citet{Esposito2016} showed that the amount of lifted dust can increase by a factor of 10 in view of high electric fields.

As far as airborne dust size distributions are concerned, there have been wind tunnel experiments where the dust fraction has been observed downstream \citep{Gillette1974, Alfaro1998}.
As dust emission is of utmost importance for life on Earth, there are also frequent field measurements from various arid regions as discussed by \citet{Mahowald2014}.

Some of these dust size distributions are shown to be in agreement with a model by \citet{Kok2011}, who considered the particle size distribution (PSD) to be the result of fragmentation of brittle dust aggregates.
Other distributions have larger deviations in comparison to that model and may better be quantified by a simple power law. 
One goal of our work is to connect the observed dust size distribution to the size distribution of the initial reservoir of grains being present if an impact ejects dust.

There are not many further analytic models to explain dust size distributions beside the brittle fragmentation model by \citet{Kok2011}. One difficulty is that experimental or field data which are compared against each other, mostly refer to data collected in well developed saltation cascades. These cascades are rather complex as many parameters such as impactor sizes, collision velocities and soil properties determine the outcome. Additionally, the effect that individual saltator impacts have on overall dust emission is difficult to evaluate under continuous saltation conditions. In that regard, we aim to provide experimental data on what PSD a limited number of only a few independent impacts generate from a well characterized soil.

We are mostly interested in dust particles with sizes of a few \textmu{}m, which can be suspended on Mars.
With grains of this size, there are a couple of studies on ejection related to planet formation. In these studies, dust aggregates and their collisions are a fundamental process \citep{Wurm2021nat}.

It might be worth noting at this point that not only the size of a free particle plays a role, but also its composition. If it is an aggregate consisting of smaller grains for example, its stability upon collisions is greatly reduced. Furthermore, as gas drag is dependent on the surface to mass ratio, aerodynamic properties are influenced by particle composition as well. However, not every non-spherical particle has to be an aggregate in this sense if its individual grains are rather cemented together.

Fragmentation of dust aggregates consisting of micron-sized grains usually creates power-law size distributions of fragments, where the number of fragments steeply decreases toward larger grain sizes \citep{Wurm2005, Deckers2014}. This is in stark contrast to the size distributions of aeolian dust mentioned above \citep{Gillette1974,Alfaro1998,Mahowald2014}. 
Another study by \citet{Krijt2014} gives a model for the size distribution of aggregate fragments. According to it, an aggregate purely consisting of 1\textmu{}m particles, would break into fragments (also aggregates) with a minimum size substantially larger than 1\textmu{}m. Following this model, even a soil containing fine 1\textmu{}m dust, would not necessarily release individual grains upon impact of a saltating particle.
\citet{Dominik1997} studied the collisions of micron-sized dust aggregates (including aggregates consisting of particles of different sizes) and quantify at which speed or impact energy individual (sub)micron-sized grains and aggregates are ejected. However, they give no size distribution and their study would not account for a wide range of particle sizes being present in the soil. 

This shows that there is not a simple model to account for all details of dust ejection in an arrangement of brittle, granular and cohesive matter. 
The outcome of saltation on dust emission depends on the soil and especially the behaviour of the cohesive 1\textmu{}m size fraction is not well understood.

With this in mind, we study the effect of individual impacts into well prepared Martian analog soil. Our major goals are to determine (1) if the most cohesive fraction of small \textmu{}m-sized dust particles can directly be released upon impact of a (saltating) sand grain, (2) if there is a generic size distribution of ejecta, and (3) if we can pin down in more detail how much mass of dust in the suspension size range can be ejected in a single impact, i.e. to give an absolute value to scale the size distribution from point 2.

\section{Experiments}

To achieve these goals, a number of different aspects have to be considered for impact experiments. First, a suitable, well-characterized Martian analog soil has to be set up. Second, an experiment has to be set up to generate impacts with a limited number of grains of suitable size into this soil. These impacts have to be independent of each other. The impact parameters (angle, speed) should be appropriate for impacts of sand grains at the saltation threshold on Mars.
Finally, these impacts and especially their microscopic ejecta have to be observed 
to allow a quantification of size and, in the best case, absolute ejecta mass or volume determinations.  

\subsection{Soil: Martian simulant}

\begin{figure}
	\includegraphics[width=\columnwidth]{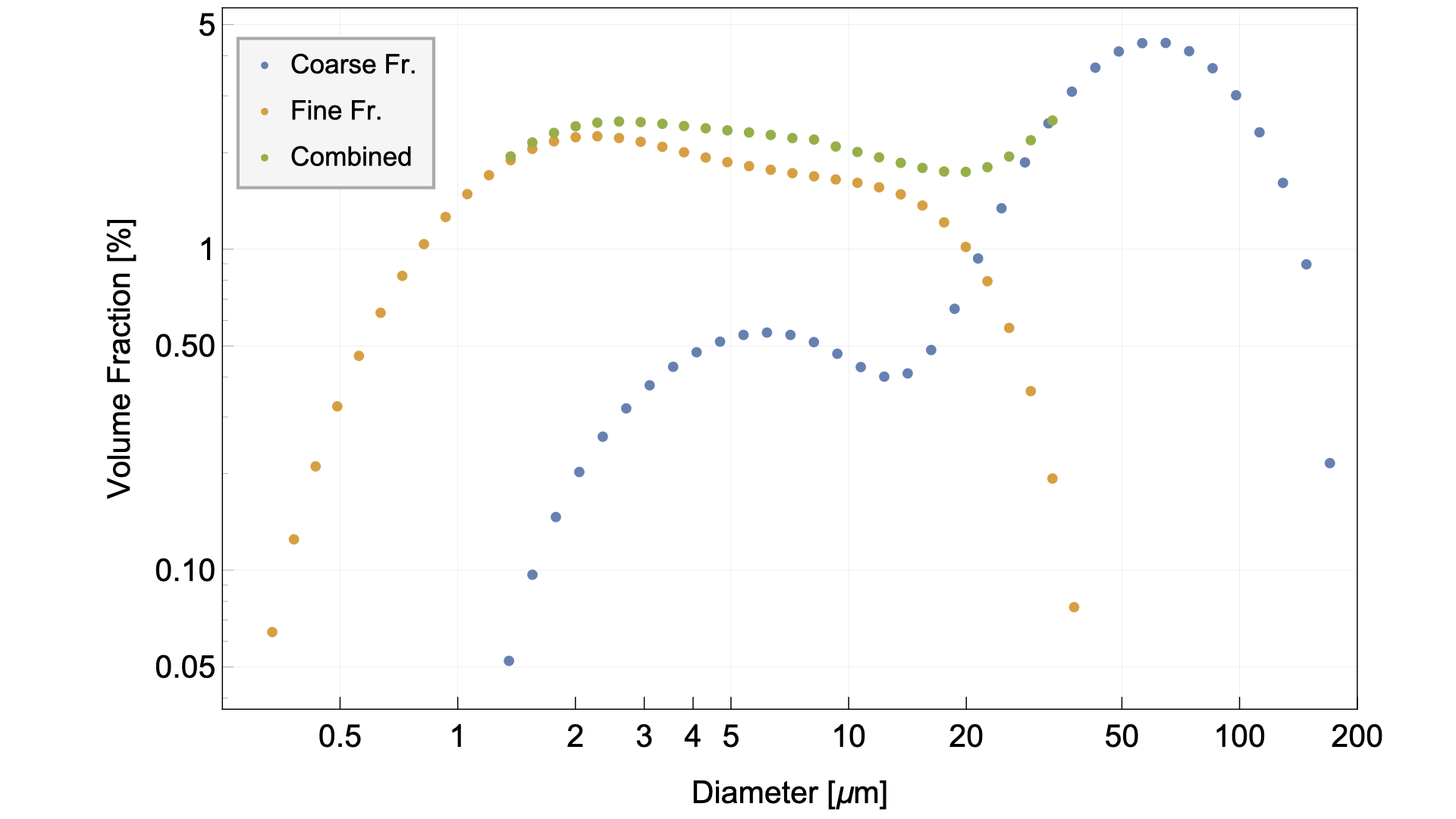}
	\caption{\label{fig.initiallarge}Volume size distributions of the MGS sample; The sample is a mix of two populations of equal total volumes.}
\end{figure}

There are several soils currently in use to serve as Martian analogs with respect to different parameters. We choose the \textit{Mars Global Simulant} (MGS) here \citep{Cannon2019} as it is based on a mineralogical composition which we consider appropriate for simulating cohesion and impact dynamics. We conditioned a sample according to the following rationale:
The top layer of Martian soil is strongly dependent on the sedimentation of airborne dust.  Ongoing sedimentation of airborne dust provides a thick top layer of \textmu{}m-sized dust. However, if saltation occurs periodically - which after all is the premise of the dust liberation studied here - there would be a continuous mixing of large saltators and dust. In an idealized case, we therefore have two important size scales: the large saltating grains and the atmospheric dust. With the measured size distributions on Mars for the large (silt to sand size) grains and general scale of saltating grains in mind, we prepared the coarse fraction of particles with sizes between 20 and 180\textmu{}m by sieving the original MGS sample. The size distribution of this coarse fraction is shown in fig. \ref{fig.initiallarge}.

To yield a fine fraction of suspendable dust  ($\sim$ 1\textmu{}m) the original MGS sample was first milled for one hour using a Retsch PM100 planetary ball mill. The milled powder was then suspended in a liquid medium and subjected to a high-performance dispersion process for a few minutes using a high-shear stirring device (Ultra-Turrax). The dispersed suspension was finally poured onto a 20\textmu{}m metal sieve for a wet sieving separation of the fine silt to clay particles. 
\begin{figure}
	\includegraphics[width=\columnwidth]{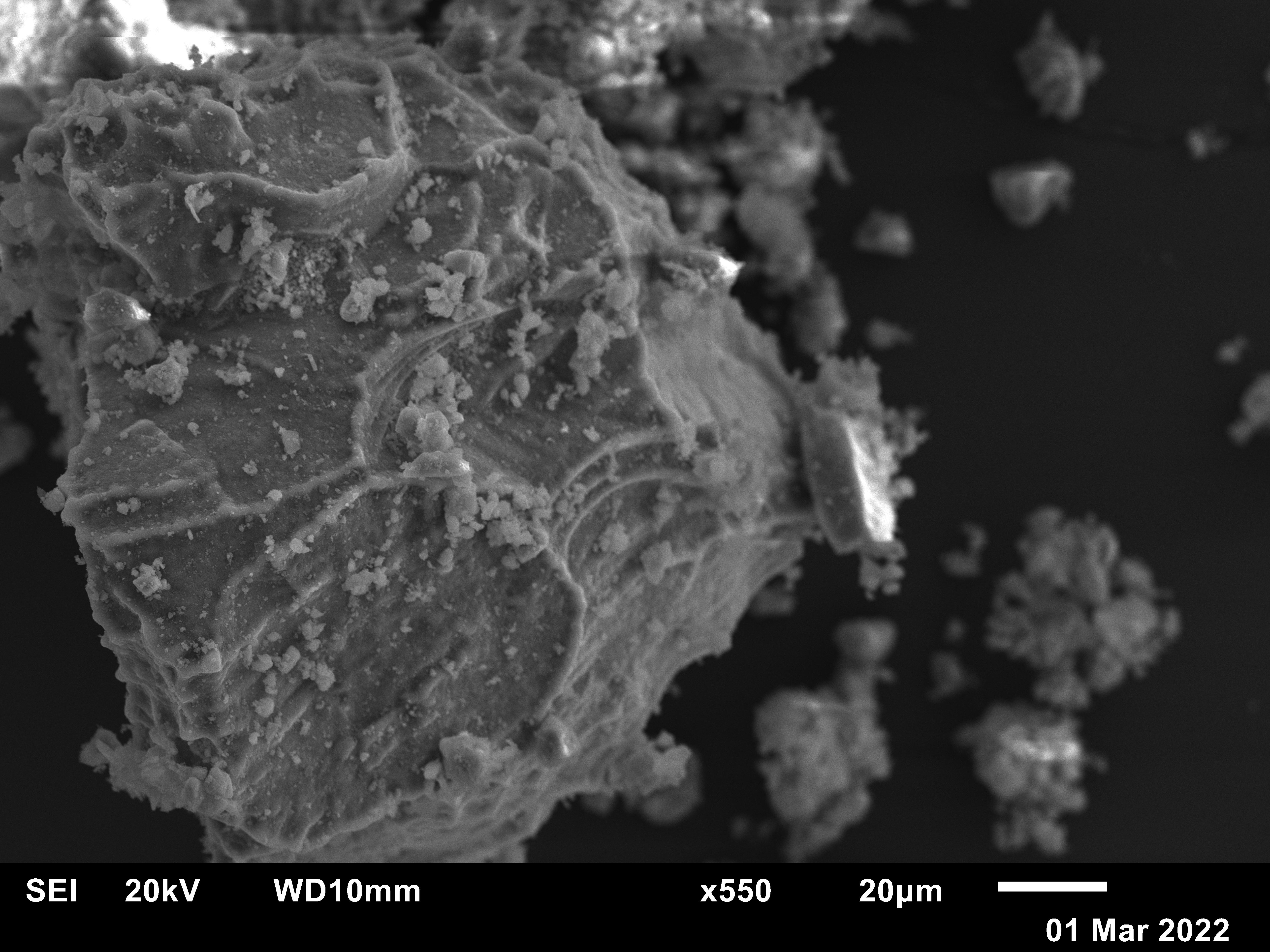}
	\caption{\label{fig.sem}SEM image of the MGS sample. Dust of the fine fraction is sticking to the surface of a particle of the coarse fraction. }
\end{figure}
This is the fraction that is important for us as the suspended dust is drawn from this reservoir. We therefore consider the volume size distribution of this fraction in comparison to our measurements. The size distribution of this fraction can also be found in fig. \ref{fig.initiallarge}.

We prepared the soil by mixing the dust fraction with the coarse 20-180\textmu{}m particle fraction. For the mixing ratio we picked 1 to 1 mass fractions. This might be somewhat arbitrary, however, if the sample is stirred afterwards we expect some self-regulating in how much small dust sticks to individual large grains of the coarse fraction, though this may be debatable. 
Certainly, the influence of the mixing ratio should be studied in the future but that is beyond the scope of this study. 
The attained sample is then used as the top layer of our dust bed. Its thickness is at least $2$mm, covering the filling material, which only consists of the coarse grain fraction. An SEM image of a particle from the coarse fraction, covered with smaller silt and clay grains can be seen in fig. \ref{fig.sem}. Even though the experiment is performed under terrestrial, rather than Martian gravity, we deem that of minor importance, as efficiency of splashing events only shows a weak dependence on gravity \citep{Duran2011}.

\subsection{Impact setup}

\begin{figure}
	\includegraphics[width=\columnwidth]{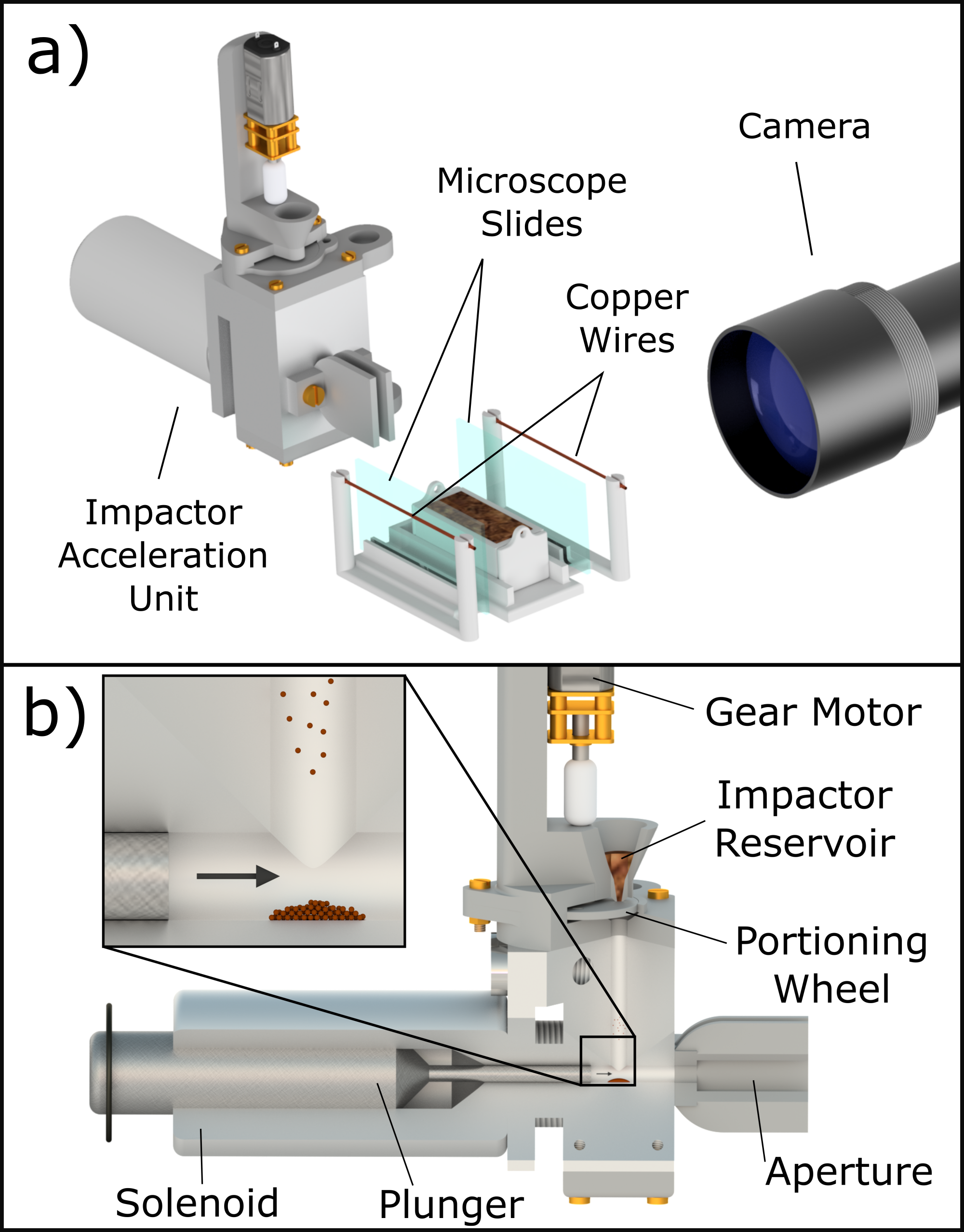}
	\caption{\label{fig.aufbau}Sketch of the impact setup; a) accelerator and impact bed with microscope slides to the side, trapping airborne dust and wires to provide an inhomogeneous electrical field. Camera on the right (distance bed-camera not to scale); b) cross section of the accelerator unit; About 20-50 impactors are released from a particle reservoir by rotating (gear motor) a portioning wheel. They fall in the path of the plunger and are launched when it is accelerated by the solenoid; the aperture guides the impactors to the soil.}
\end{figure}

The soil prepared this way is used for impact experiments. A sketch of the setup is shown in fig. \ref{fig.aufbau}. 

\begin{figure}
	\includegraphics[width=\columnwidth]{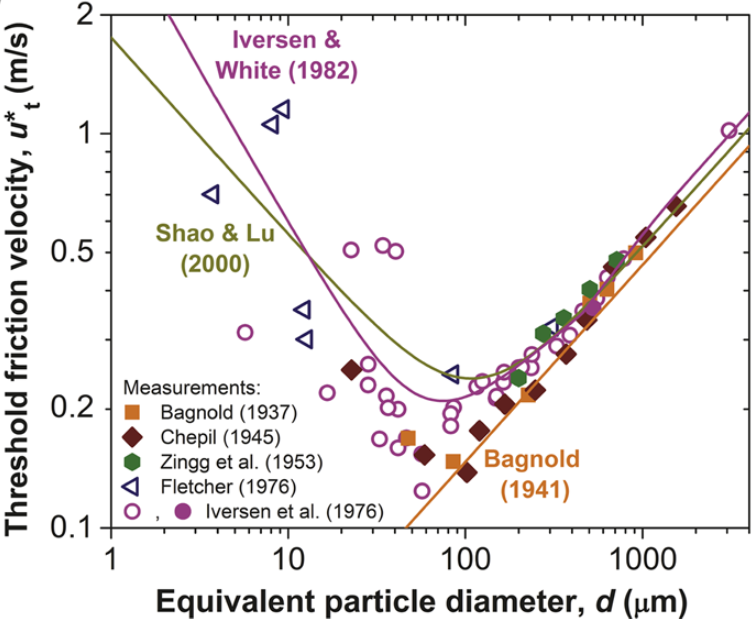}
	\caption{\label{Threshold}Curves of Threshold friction velocities depending on grain size for different models as well as experimental results. (taken from \citet{Neakrase2016})}
\end{figure}

The impactors are drawn from the same material (MGS) as the dust and coarse fractions. We used a size fraction between 180 to 250\textmu{}m for the first setup. Smaller sized particles are not only more difficult to ration, but also pose a problem for any moving parts of the dispensing mechanism. According to the threshold curve in fig. \ref{Threshold}, particles of 100\textmu{}m would have been most desirable; however, even though 100\textmu{}m particles are the fraction with lowest threshold velocity according to the plot, larger particles certainly are moved by wind as well, as observations have shown \citep{Greeley2001}. Additionally, fig. \ref{Threshold} shows that the threshold windspeed for 250\textmu{}m sand particles is still lower than eg. for 20\textmu{}m grains. A Study performed by \citet{Swann2020} even suggests, that 200\textmu{m} particles require equal or even lower threshold wind velocities than 100\textmu{}m particles to be moved under Martian conditions. Thus, we are confident that we are within a reasonable size range to simulate saltators on the Martian surface.

These impactors are accelerated by a plunger and impact the soil at angles of $18.8^{\circ} \pm 2.5^{\circ}$. This value fits well into the range suggested by several studies \citep{Bagnold1941, Chepil1945, White&Schulz1977, Jensen&Sorensen1986}. 
Impact velocities are in the range of $1.04 \pm 0.2$m/s and thus in agreement to the shear velocity at the threshold of saltation \citep{Swann2020}. Due to the grain size being slightly larger than particles at the saltation threshold, however, impact energies are somewhat higher. This should not have too big of an impact, though. A study conducted by \citet{Bogdan2020} showed, that there is no correlation between impact energy and energy dissipated to the ejected particles for small ($\pm 50\%$) variation in impact energy. Even though their setup was slightly different from the one used in this study, as impacts occured in a $90^{\circ}$ angle and the ratio of impactor to target particle size was smaller, we still believe there to be enough parallels for their findings to be somewhat relevant to our case.

\begin{figure}
\includegraphics[width=\columnwidth]{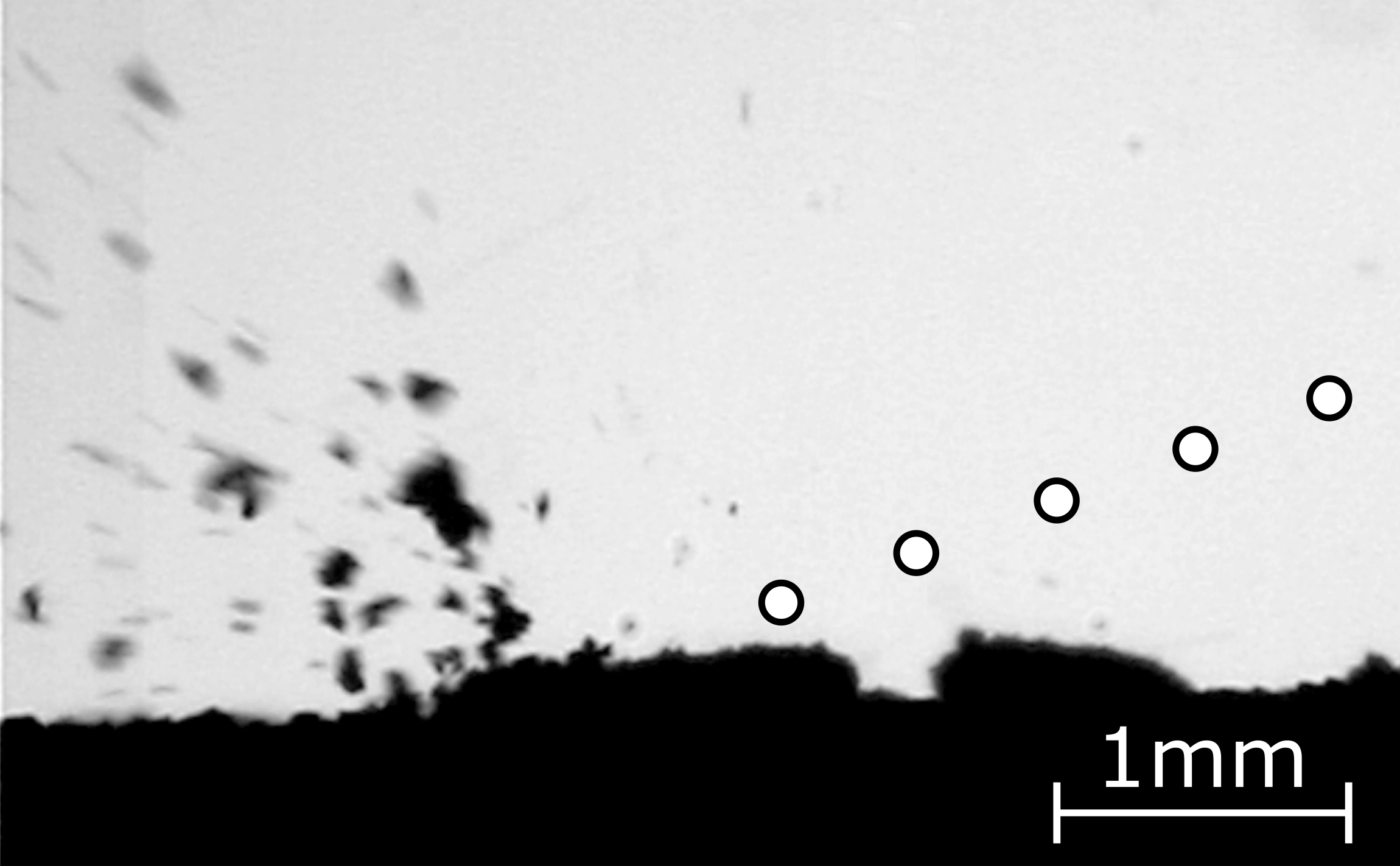}
	\caption{\label{fig.impact}Image of an impact, showing the impactor as white circle as well as a cloud of ejecta.}
\end{figure}

Each run, about 20-50 particles impact the model soil. The impacts are well separated, so that we do not expect one impact to influence another \citep{Bogdan2020}. We therefore see the sum of the outcome of a small number of individual impacts.
An exemplary image of such impact is shown in fig. \ref{fig.impact} including the splash and the impactor trajectory, implied by a white sphere. The footage of the impacts, taken with a NAC MEMRECAM HX-3 at 5000 fps, can be used to analyze impact angles and speeds. The camera was mounted 15cm to the side of the dust bed with the lens being in line with the bed's surface (fig. \ref{fig.aufbau}).
The important small dust fraction is not visible with this method.
Instead, it is captured on two microscope slides placed next to the soil as seen in the sketch in fig. \ref{fig.aufbau}. The capture efficiency is increased by electrostatic attraction as two electrodes are placed behind each slide with a voltage of about $1.6 \rm kV$ applied between them.

As the micrometer dust is susceptible to gas drag - this is why it is suspended even at hPa Martian pressure in the first place - it would easily be missed in the experiments if it couples to a convective flow.
Therefore, the experiment is placed within a vacuum chamber and operated at a pressure of $2 \times 10^{-2} \rm mbar$. 

The microscope images taken with an AxioCam ICc 1 mounted on a Zeiss AxioImager.M2, are our main raw data.
As the contrast is high in these images, we segmented each image by setting a simple threshold brightness. There is no large sensitivity to this threshold, as the small dust grains are well focused and thus have a clear outline on the images.

\section{Results}

\begin{figure}
\includegraphics[width=\columnwidth]{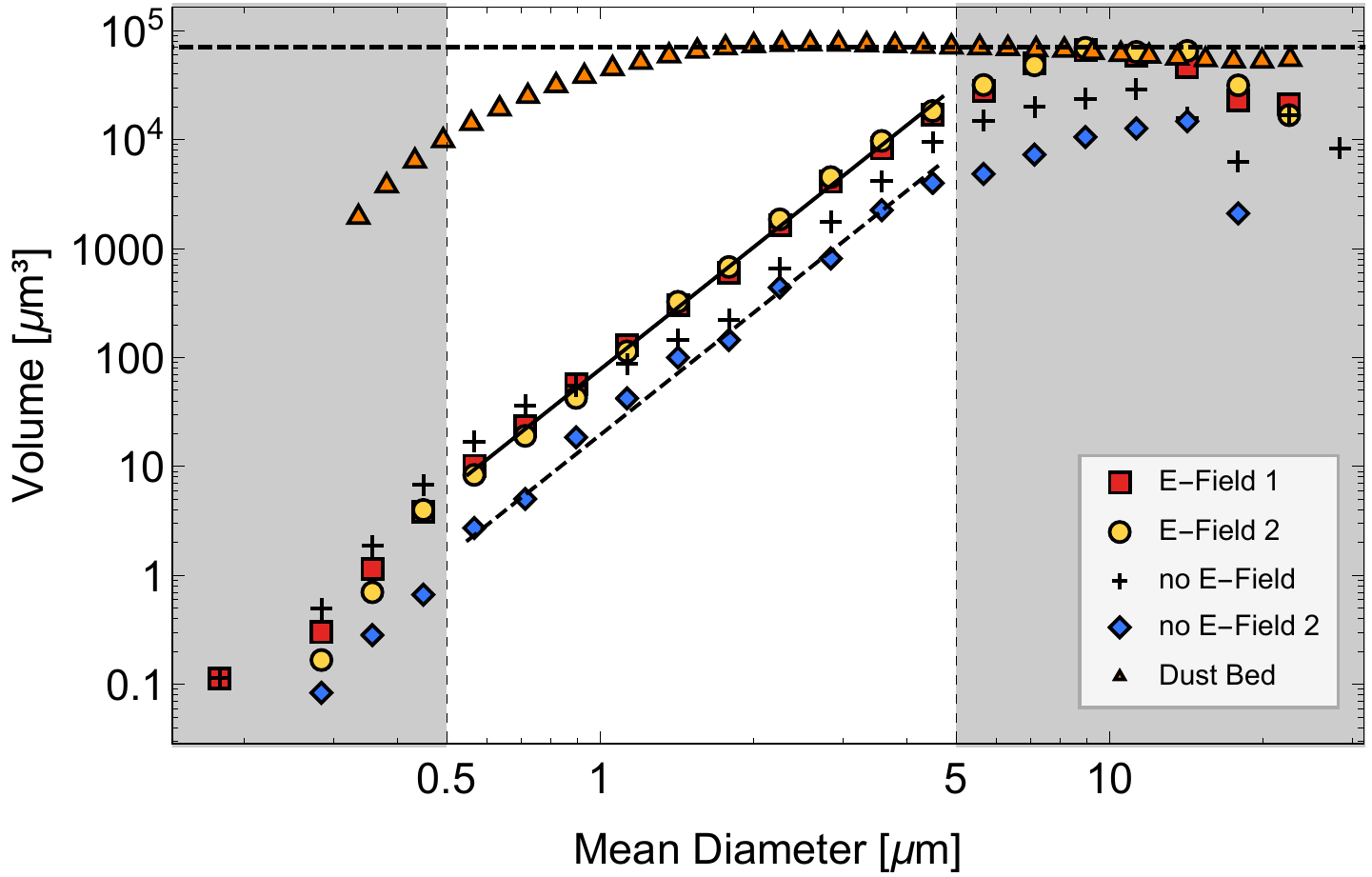}
	\caption{\label{fig.smallgrains}Volume size distributions of ejected dust grains for the 2 runs with electric field applied and one run without. Overplotted is a power law fitted to the data (black solid line).  
	The same fit, shifted by half an order of magnitude, is overplotted for the experiment without field (black dashed line).
	Also seen is the initial size distribution used in the sample (dust bed). It is approximated with a dashed horizontal line.
	}
\end{figure}

In total the data analyzed and presented here are from two runs of impact experiments with electrical field applied and two without electrical field for comparison. 
Each run consists of 7 individual launches, firing between 20 to 50 impactors each time. Afterwards, we analyzed 600 images from the slides on both sides.
For each identified particle, we measured the cross section and derived the diameter $s$ of an equivalent area sphere. In fig. \ref{fig.smallgrains}, we show the retrieved dust size distributions together with the initial size distribution of the soil. 

For the two runs with electric field turned on, we count a total of 4738 and 4921 particles, respectively. For the two runs without field the counts are 1115 and 3018. 
The increase in captured dust with an electrical field being present shows that there must be some sort of charge or dipole moment present on individual dust grains after the impact. If that was inherent or due to triboelectrical processes during the impact cannot be determined. In view of electrostatic lifting mechansims mentioned above, it cannot be ruled out that electrostatics add a lifting force here.

The electrical field also seems to remove some variability in the size distributions. This can be explained by variations of trajectories of ejected material. Due to the electric field, many particles with trajectories not aimed at the slides are being captured as well, thus, homogenizing the number of detected grains. 

Overall, we see a significant decrease (40-80\%) in particle count when doing experiments without the electrical field, which might be due to a different lifting efficiency or capture efficiency.
Most important though, as seen in fig. \ref{fig.smallgrains} the slope of the size distribution without field is the same as with field for the fraction $<$5\textmu{}m. This implies that electrostatic attraction does not induce a size bias in that range. Above 5\textmu{}m the size distributions with and without field show less similarity.

Overall, the data has a lower cut-off due to the microscope resolution but also due to the smallest grain size present in the sample. Due to the resolution, we consider the data below 0.5\textmu{}m to be only of a qualitative nature, showing that also dust of this small size is ejected. In fact, all particle sizes present in the initial size distribution are lifted. We consider it an important finding that impact emission really includes the smallest grains which are very cohesive and are not easily removed by direct wind lifting.

The detection of larger particles holds a bias as well. They are less susceptible to the electrical field due to their higher mass. Additionally, with increasing size the probability of sticking to the glass slides greatly decreases, lowering the number of emitted particles that are being captured.
We therefore only consider the data below 5\textmu{}m to give a quantitatively consistent size distribution. This is not a limitation to our task, as airborne Martian dust is restricted to even smaller grains under fair weather conditions.

In the unbiased size range from $s = 0.5-5$\textmu{}m the volume size distribution can well be approximated by a power law
\begin{equation}
    \frac{dV}{d \left( \log \left( \frac{s}{\mathrm{\mu m}}\right) \right)} = a \cdot \left( \frac{s}{\mathrm{\mu m}} \right)^b
    \label{eq.powerlaw}
\end{equation}
Consistent power laws are found for the two experiment runs analyzed in detail (with electric field). We get $a=77.2 \rm \mu m^3 \pm 4.5 \, \rm \mu m^3$ and $b=3.72 \pm 0.1$, which are the averages of the two values and the range measured. 
We do not have a physical model, and as such we cannot assess whether these results are within expectations. However, in order to give a simple analytic description of the size distribution, we take this as an empirical finding that helps us in quantifying dust fluxes.

It also has to be noted that the size distribution of the two runs are not scaled in any way so the data show that not only the power law matches well but also the absolute values are very robust with a variation for a given size bin of typically less than  10 \% in the size range between  0.5 to 5\textmu{}m. This implies that the average absolute value of ejected dust, in principle, can be quantified on this level.

\subsection{Total volumes}

While the size distribution is somewhat uncertain for sizes of 1\textmu{}m or less, as outlined above, this is of minor importance for the total volume or mass of grains ejected up to a certain size, as the volume size distribution increases rather steeply with size. 
If we choose a hard upper cut-off diameter $s_{cut}$, motivated by larger dust not being capable of being suspended, this total volume can be estimated by integration of eq. \ref{eq.powerlaw} and gives
\begin{equation}
    V_{capt} = \frac{a}{b} \cdot \left( \frac{s_{cut}}{\mathrm{\mu m}} \right)^{b}
    \label{eq.mass}
\end{equation}

This is an absolute value measured but only gives the total volume of dust actually captured. It thus only refers to the sample area of the slides that has been imaged and analyzed. To get an estimate for the total amount of dust, we have to introduce a correction factor $f_{capt}$ which relates the true volume of ejected dust $V_{true}$ up to $s_{cut}$ to the captured volume $V_{capt}$
\begin{equation}
    V_{true} = \frac{V_{capt}}{N \cdot f_{capt}} 
    \label{eq.volume}
\end{equation}
Here, $N$ is the number of impactors per run. It varies between 20 and 50 due to the dispension mechanisms inaccuracy. On average, a number of 27$\pm$12.7 particles impacts the dust bed every launch, summing up to $N = 189 \pm$38 over all seven runs of the experiment. 

\subsubsection{Estimate of $f_{capt}$}

As fig. \ref{fig.smallgrains} shows, the applied electrical field strongly increases the amount of captured dust. It also keeps the power of the size distribution the same. As electric forces on tribocharged grains are size dependent, but as we do not see a size fractionation in the discussed size range with electric field turned on, this suggests that essentially all free dust grains are attracted by the electrodes and detected by the slides. This is certainly a simplification but the easiest one to chose due to lack of information on the grains.

We obtained our data by taking a series of images of the dust trapped on the microscope slides. 
Overall, only $\sim 1 \%$ of each slide was imaged, as about 36000 individual images would be necessary to cover the entirety of a slide. However, we only acquire 300 images per slide.
Not every region of the microscope slides is covered with dust. As those areas have no quantitative value for us, there aren´t any images taken of them, as would be when just picking 300 random locations on the slide. Thus, the area of the 300 images taken corresponds to a slightly larger area that has been scanned for dust. Taking this biased sampling into account, a reasonable factor from only analyzing parts of the slide would then be $f_{capt} = 1/100$.
Certainly, this is only a rough estimate. However as a first benchmark, it might suffice.

\subsection{Ejection probability}

We want to relate the ejected as well as the detected mass (volume) and size distribution to the initial size distribution of the soil. 
We show the initial arbitrarily scaled size distribution in fig. \ref{fig.smallgrains}. 

The size distribution of the initial sample has been determined by light scattering with a Mastersizer 3000, thus, possibly resulting in a small difference in the definition of size with respect to the microscopic size determination. Both kind of distributions might therefore be shifted slightly relative to each other.
Fortunately though, the original volume size distribution is essentially flat in the studied range, noting some difference at grain sizes of 1\textmu{}m and below. Compared to the steep power law of ejected dust we consider this to be of minor importance. To first order, we can therefore approximate the initial size distribution as constant. 

Relatively larger particles imaged on microscope slides could be aggregates of smaller sized grains but this is not discernible in our data. 
Due to their higher porosity or surface over mass ratio, aggregates are more susceptible to electrical fields and also picked up by the wind more easily. In that regard it would be of interest to know whether or not and, if, what fraction of the particles we capture are  aggregates. While the optical analysis methods of this study are not sufficient to bring deeper insight into this matter, we are currently setting up another experiment with the goal of studying just that matter. That being said, concerning this study we settle with just analyzing the sizes of captured particles and set aside any other properties these particles might have.
As such, with the flat initial distribution, the power law found for the ejected dust can be assumed to be a direct probability density $\rho$ of how much dust of a given size is ejected, or
\begin{equation}
    \rho = \frac{dV_{true}/d log s}{dV_{init} / d log s} \sim s^b.
\end{equation}

\subsection{Absolute emitted dust masses}

The probability distribution quantifies the size distribution of ejected grains and may be adapted to more complex initial size distributions. Besides the size distribution, the absolute masses or volumes ejected are important. Our data provide some insight on this as well. For the given parameters of (1) flat and slow impacts of $\sim$ 1 m/s, (2) flat initial fine dust distribution, (3) a mass ratio of 1 to 1 of fine to coarse grains, and (4) an average impactor size of $\sim 200$\textmu{}m (volume of $V_{imp}=4 \cdot 10^{6} \rm \mu m^3$), we find an absolute ejected dust mass according to eq. \ref{eq.powerlaw} and \ref{eq.mass}.

Therefore, in our special setting, a ratio of
\begin{equation}
    R = \frac{V_{true}}{V_{impactor}} = c \cdot \left( \frac{s_{cut}}{1\mathrm{\mu m}}\right)^b
\end{equation}
is ejected as dust with $c = a/(b \cdot N \cdot f_{capt} \cdot V_{imp})= 3 \cdot 10^{-6}$
$(a = 77 \mathrm{\mu m^3}, b = 3.72, N = 189, V_{imp} = 4 \cdot 10^6 \mathrm{\mu m^3})$.
That e.g. gives an ejected mass of $R =  2 \cdot 10^{-4}$ times the impactor mass ejected in dust below 3\textmu{}m per single impact.

\section{Caveats}

We note that these results and mass estimates are based on the kind of soil we prepared in the laboratory. For a thick dust layer, fluffy or compact or weathered soils with quite different particle sizes and sticking properties between grains, this might and will be different.

The results are also tied to the specific impact velocity we used. While this was motivated from other experiments on wind driven matter at low gravity just beyond the threshold, other conditions are thinkable. Small variations in impact velocity might not change the results much \citep{Bogdan2020}. However, for strong increase there will be an increase in ejecta mass \citep{Colwell2008}. How this would change the size distribution of the fine dust fraction, we cannot say at this stage.

Our upper cut-off of 5 \textmu{}m is the largest size for which we would argue that our size distribution follows the power law given. Everything beyond could be extrapolated but with some uncertainty. If these particles on the order of 10 \textmu{}m would turn out to be high porosity aggregates with low sedimentation speeds, they might become important.

The initial size distribution shows some deviation below 1 \textmu{}m from the flat distribution we assumed. While this small size fraction is small in volume compared to the volume fraction of larger grains due to the steep power law and while the different techniques of size measurements might account for this, it should be kept in mind that our ejection probability might be less accurate below 1 \textmu{}m.

\section{Conclusions}

In laboratory experiments, we quantified the amount and size of grains ejected after individual impacts of sand-sized grains into a Martian analog soil at low pressure.
We find that the probability of releasing dust grains of a given size from the soil in the range from 0.5\textmu{}m to 5\textmu{}m is a steep power law with power $3.72$. Thus, even though the probability of ejection increases with grain size in this range, there is no hard cut off. Meaning, even the smallest sized dust of this fraction is released upon impact. An estimate using the given parameters shows: a fraction of $2 \cdot 10^{-4}$ of an impactor's mass is released in suspendable dust below 3\textmu{}m in an impact. We conclude that saltation is a source of dust grains in the Martian atmosphere.

\section{Acknowledgements}

This project has received funding from the European
Union’s Horizon 2020 research and innovation
program under grant agreement No 101004052. We appreciate two very helpful reviews by two anonymous referees.



\bibliographystyle{aasjournal}
\bibliography{bibbi}



%
%


\end{document}